\documentclass[showpacs,aps,preprintnumbers,amsmath,amssymb,letter,twocolumn]{revtex4}
 \usepackage{graphicx}
\usepackage{dcolumn}
\usepackage{bm}
 \usepackage{amssymb}
 \usepackage{amsmath}
 \usepackage{mathalfa}
 \usepackage{color}
 \usepackage{natbib}
 \usepackage{url}
 \usepackage{multirow}
\usepackage{subcaption}
\usepackage{caption}
\begin{document}

\title{Transport Barrier generation \\ at the interface of regions with different zonal flows dynamics}

\author{C. Norscini}
 \email{claudia.norscini@cea.fr}
  \author{Ph.Ghendrih}
 \author{T. Cartier-Michaud}%
 \author{G. Dif-Pradalier}%
  \author{X.Garbet}%
  \author{N.Nace}%
 \author{Y. Sarazin}%
  \author{P. Tamain}%
\affiliation{ CEA, IRFM, F-$13108$ St. Paul-lez-Durance cedex, France} 
\date{\today}

\begin{abstract}
A novel and generic understanding of spontaneous generation of transport barriers and zonation regimes in turbulent self-organization is presented. It associates the barrier onset to the development of a spectral gap between large scale flows and turbulence modes leading to a zonation regime. A robust barrier builds-up at the interface of such a region and a neighboring one with reduced zonal flow generation. This more complex and generic transition paradigm could fit the numerous and sometimes conflicting observations as in fusion plasma experiments. Barrier relaxation by bursts of turbulence regenerate the zonal flows that are eroded by viscous (collisional-like) damping. The duration of the quiescent phase between the quasi-periodic relaxation events is governed by this damping process, hence the barrier collision frequency for fusion plasmas.
\end{abstract}
\maketitle
Turbulence self-organization plays a major role in transport properties within stratified media in geophysics \cite{Maximenko2005}, astrophysics \cite{Vasavada2005} as well as laboratory plasma dedicated to magnetic fusion research \cite{Burrell1997,Fujisawa2009}.
An outstanding mechanism is the transition from a fully turbulent to the so-called zonation regime \cite{Galperin2008}, where large scale anisotropic flows, the zonal flows, appear to undergo condensation to a regular pattern \cite{Tobias2013,Parker2013}. These regimes are investigated via reduced 2D models of the Hagesawa-Mima type with prescribed energy injection in the spectra of otherwise decaying turbulence. They are regarded as a possible mechanism for the generation and maintenance of stable systems of alternating zonal jets as observed in the deep terrestrial oceans and in the atmosphere of giant planets \cite{Galperin2010}. 
In the fusion plasma framework,  self-organized zonal flows are considered to govern transport barriers \cite{Dif-Pradalier2015,Diamond2005}. To study such relation, particle (or heat) transport has to be taken in account in the 2D model.\\
Without loss of generality, we address such problem via a model comparable to the Rayleigh-B\'enard (RB) instability, where self-consistent turbulence drives both the transport properties and the energy injection into the turbulent modes via the interchange (buoyancy) instability.
We report here the formation of a transport barrier, or pedestal, localized at the interface between two regions with different zonal flow damping capability. For magnetic fusion plasma, that correspond to the edge region, where the magnetic surfaces are closed, and the Scrape-Off Layer (SOL) region \cite{Sarazin1998}, where the magnetic surfaces intersect wall components. This interface is referred to as the separatrix.
Of interest here is the role of large scale damping in controlling the transport regime \cite{Vallis1993,Sukoriansky2007} and consequently the transport barrier.\\

 The 2D fluid model of the interchange instability \cite{Garbet1991,Sarazin1998}, is similar to several models addressing turbulence generic properties. It is simplified to only retain the key features required for the barrier generation. It stands for particle and charge conservation equations in an isothermal plasma in the cold ion limit:
\begin{subequations}
\begin{eqnarray}
\partial_t n + \left[\phi,n\right]  \!&-&\! D_\perp \Delta_\perp n = S - \Gamma
\label{density_equation} \\
\partial_t W + \left[\phi,W\right] \!&-&\! \nu_\perp \Delta_\perp W+ ~g\partial_y n = 
 J
\label{vorticity_equation}
\end{eqnarray}
\end{subequations}
where $n$ is the particle density and $W$ the vorticity, $W=\Delta_\perp\Phi$. Time and space are normalized by the inverse of the cyclotron frequency and the Larmor radius respectively. Space is reduced to 2D transverse to the magnetic field assuming symmetry along the magnetic field. The $y$ coordinate is an angle, the poloidal angle for magnetic confinement, the $x$ coordinate in the radial direction that extends from the source $S$, localized in the edge, to the sink $\Gamma$ in the SOL. The separatrix is defined at a given $x_{Sep}$ position.
 Constraints due to the physics along the magnetic field govern volumetric source terms: the parallel divergence of the electric current $J$, the particle sink $\Gamma$ specific of the SOL region. Convective turbulent transport $[\Phi,f]=\partial_x (f(-\partial_y\Phi))+\partial_y (f(\partial_x\Phi))$ competes with weak diffusive transport with coefficient $D$ for the particles and $\nu$ for the vorticity.  The buoyancy $g-$force governs the plasma interchange instability and thus the self consistently energy injection into the turbulent spectrum. $\phi$ is then the electric potential and $W$ is related to the polarization charge. For standard fluids $\phi$ is the current function. In the RB model $n$ stands for the temperature variation.
The change in field line properties at the separatrix  is taken into account by a mask function $\chi(x)$, $\chi(x > x_{Sep})=0$ and $\chi(x \le x_{Sep})=1$ such that $J=\sigma_\phi(\phi-\chi(x)<\phi>_y)$,where $\sigma_\phi$ is the normalized conductivity and $<f>_y$ is the $y-$average of $f$, similarly $\Gamma=(1-\chi(x))\sigma_n n$ where $\sigma_n$ is the particle lifetime in the SOL. 
The edge and SOL difference is twofold. First the particle loss is localized in the SOL region. This has rather little effect on the turbulence but organizes the overall stratification of the system in the x direction. 
Second, the change in J modifies the evolution equation for the zonal flow $V_z=\partial_x<\phi>_y$. The latter is governed by a balance between the non linear Reynolds stress source, and the loss terms: viscous damping at small scales, sink term $<J>_y$ at large scales. The edge constraint $<J>_y=0$ favors large scale zonal flow structures. Conversely, in the SOL, the current loss $<J>_y=\sigma_\phi<\phi>_y$, here linearized for simplicity, damps these zonal flow structures.
We investigate near marginal cases for a prescribed density gradient length $1/L_n$. The linear growth rate is given by:
\begin{subequations}
	\label{eq: growth rate and Rhine scale}
\begin{eqnarray}
\gamma_m =  \frac{\left( \nu\sigma_\phi\right) ^{1/2}}{L(K)} \left( \frac{L_R}{L_n}\frac{k_y^2}{k^2} - \frac{1}{S_c}\left(K^4 +1\right)\right) 
\label{eq: growth rate}\\
L_R = \frac{g}{\nu\sigma_\phi}~~;~~L(K) = (1+1/S_c) K^2 + 1/K^2
\label{def: Rhine scale}
\vspace*{-.1cm}
\end{eqnarray}
\end{subequations}
where $K = k /\bar{k}$, $\bar{k}^4 = \sigma_\phi / \nu$ and $S_c =\nu / D$ is the Schmidt number. This expression yields three key aspects of the near marginal regime, the order of magnitude of $\gamma_m$, $(\nu\sigma_\phi) ^{1/2}$, the threshold depending on the Rhine scale $L_R$ \cite{Vallis1993,Galperin2008} such that $L_R / L_n$ must exceed a given function of $k_x$ and $k_y$, a localization function $L(K)$ that favors $K= (1+1/S_c)^{-1/4}$ as most unstable mode. The latter effect is governed by the balance between the homogeneous damping rates, at small scale due to diffusion $\approx \nu k^2$, at large scale due to parallel currents $\approx \sigma_\phi/ k^2$. The threshold favors the modes with small values of $k_x$, excluding $k_x=0$ to ensure momentum conservation by the zonal flow structure.\\

\begin{figure}[h]
  \begin{subfigure}{0.22\textwidth}
  		\includegraphics[width=\linewidth]{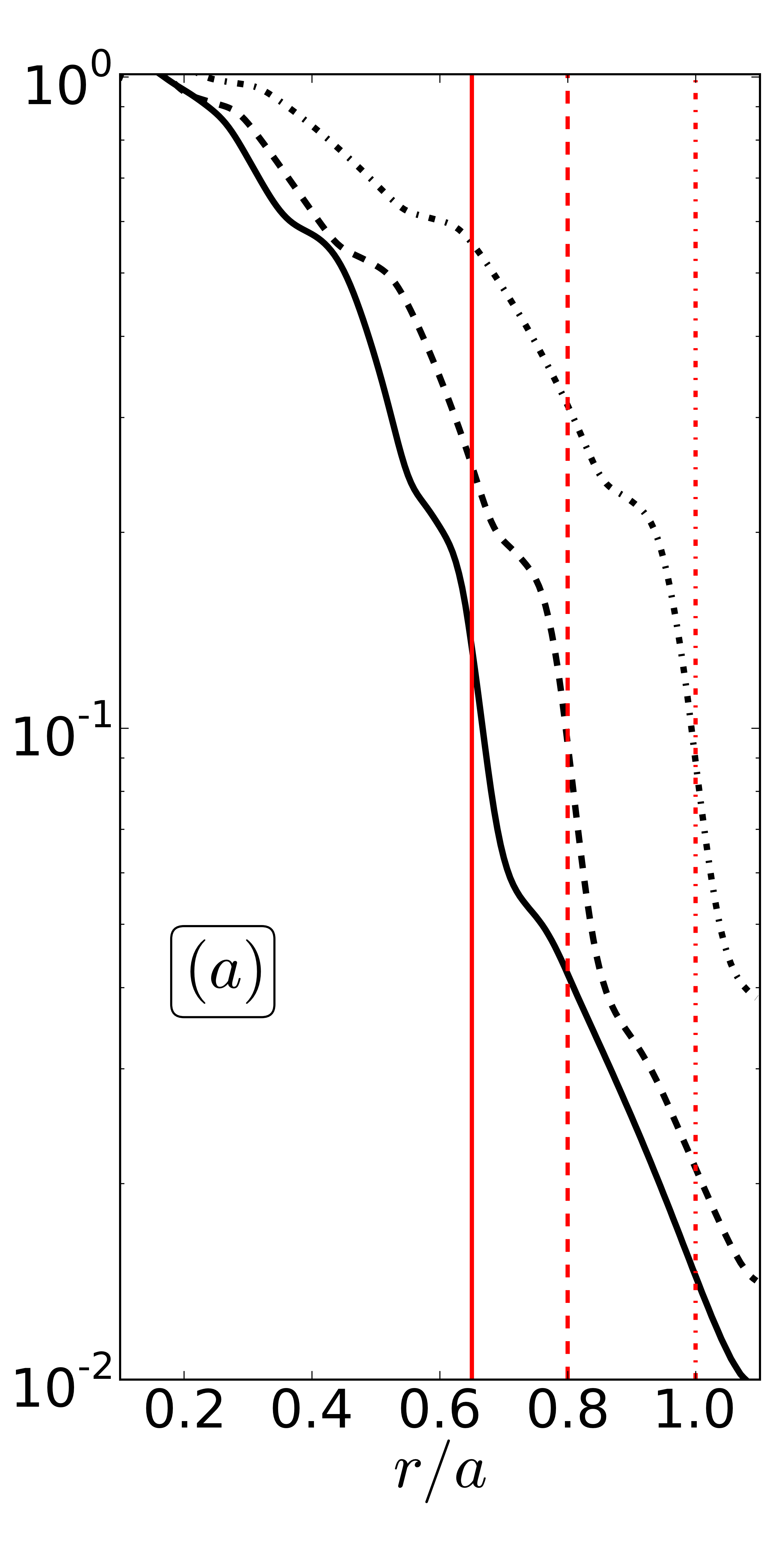}
  \caption{}
\label{subfig:fig1_a}
\end{subfigure}
 \begin{minipage}{0.2\textwidth}
\begin{subfigure}{\linewidth}
		\includegraphics[width=\linewidth]{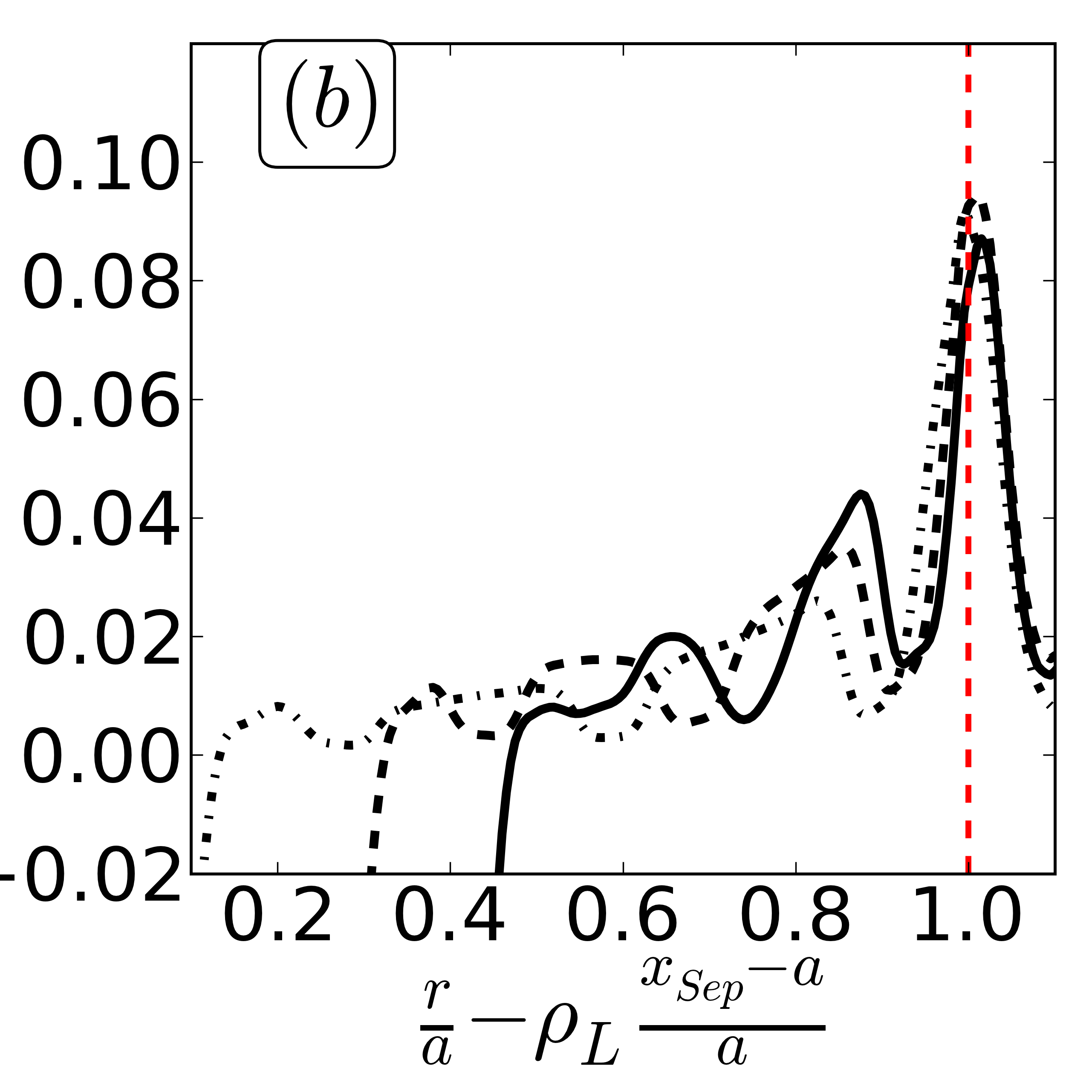}
 \caption{}
\label{subfig:fig1_b}
 \vspace*{-.7cm}

\end{subfigure}
 \begin{subfigure}{\linewidth}
		\includegraphics[width=\linewidth]{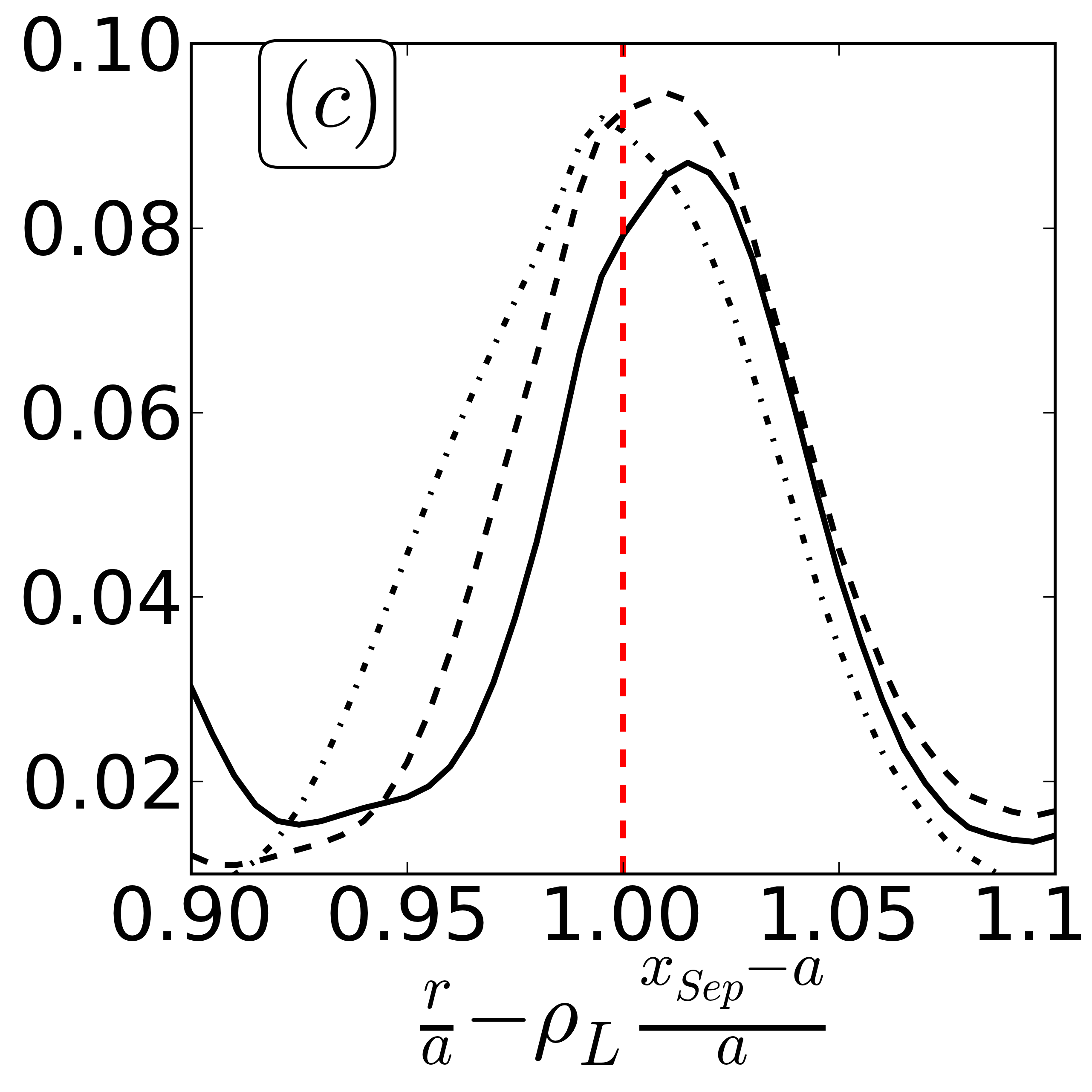}
  \caption{}
\label{subfig:fig1_c}
\end{subfigure}
 \end{minipage}
\vspace*{-.8cm}
  \caption{(\protect\subref{subfig:fig1_a}) Density pedestal for different positions of the separatrix (see text), (\protect\subref{subfig:fig1_b}) density gradient profiles with radial shift, (\protect\subref{subfig:fig1_c}) zoom on the pedestal region}
  \label{fig:fig1}
\vspace*{-.3cm}

\end{figure}

SOL-transport with the present model is comparable to that previously reported when only the SOL region was addressed \cite{Sarazin1998, Ghendrih2012}. It is characterized by avalanche transport, hence ballistic propagation of fronts and holes \cite{Sarazin1998, Ghendrih2012}, the so-called 'blobs' routinely observed in experiments. Density and potential fluctuations are large \cite{Sarazin1998, Ghendrih2012} but the mean value of the latter weakly departs from the equilibrium value, namely the plasma floating potential set at zero for convenience in this isothermal case. 
Conversely, transport in the edge region appears to be controlled by zonal flows. These generate transport barriers (TBs), where the transverse turbulent avalanches are damped and therefore where diffusive transport governs a larger fraction of the particle outflux.
At the edge and SOL interface a barrier is readily observed, fig.\ref{fig:fig1}(\protect\subref{subfig:fig1_a}). 
Three different average density profiles are compared on fig.\ref{fig:fig1}(\protect\subref{subfig:fig1_a})
where only the position of the separatrix $x_{Sep}$ is modified in the simulations, i.e. $x_{Sep}=0.6 ~x_{a}$ (continuous line), $x_{Sep}=0.8 ~x_{a}$ (dashed line) and $x_{Sep} = x_{a}$, dash-dot line). One readily observes 3 key features: the density e-folding length in the SOL is unchanged,
 the pedestal region with increased density gradient at the separatrix extends both in the edge and in the SOL region, other regions in the density profile exhibits corrugations (enhanced gradients) which drive staircase like profiles \cite{Dif-Pradalier2010,Dif-Pradalier2015}.

\begin{figure}[h]
\vspace*{-.8cm}
\begin{minipage}{0.2\textwidth}

\begin{subfigure}{\linewidth}
    \caption{}
		\includegraphics[width=\linewidth]{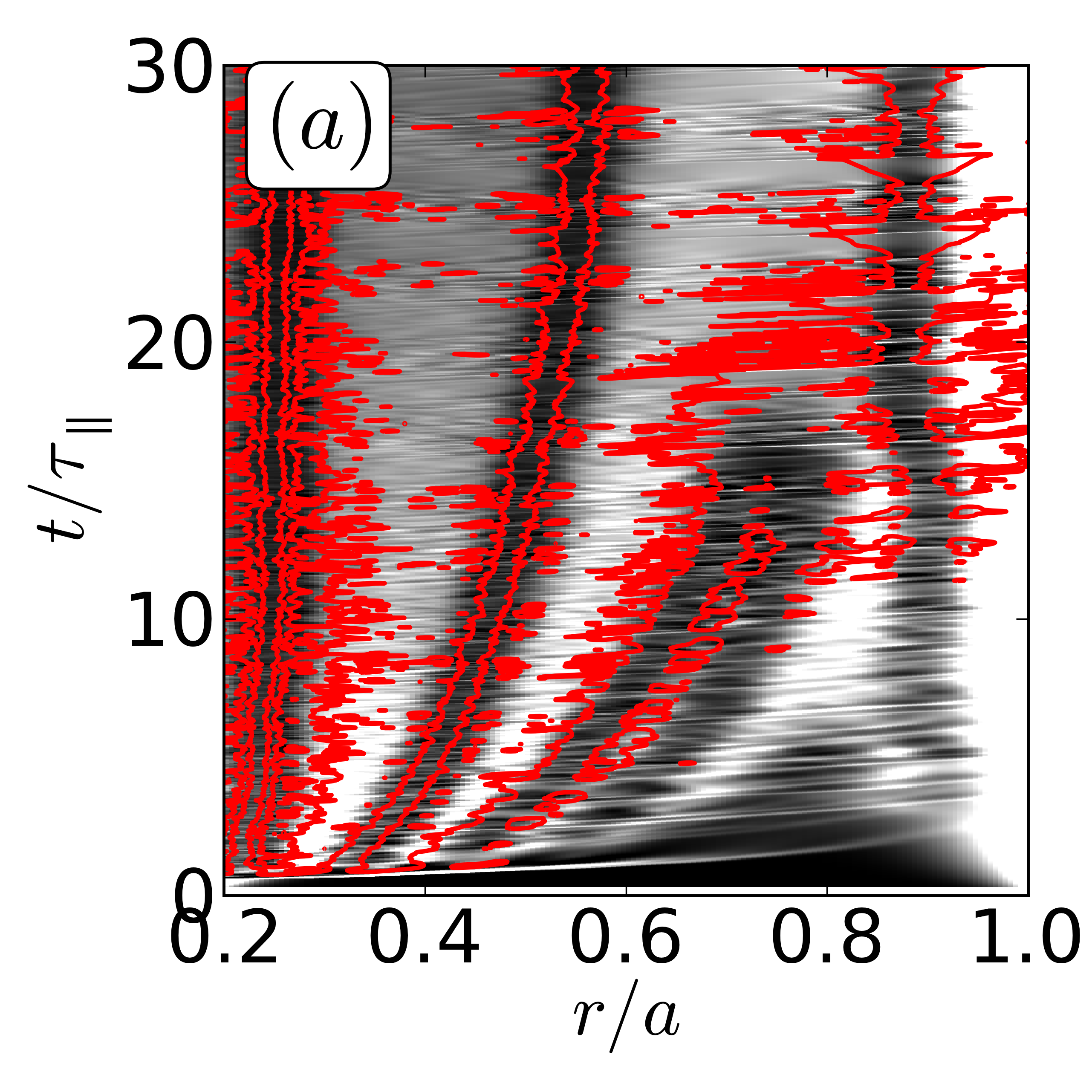}
	\label{subfig:fig2_a}
\end{subfigure}
\vspace*{-1.1cm}

\begin{subfigure}{\linewidth}
    \caption{}
		\includegraphics[width=\linewidth]{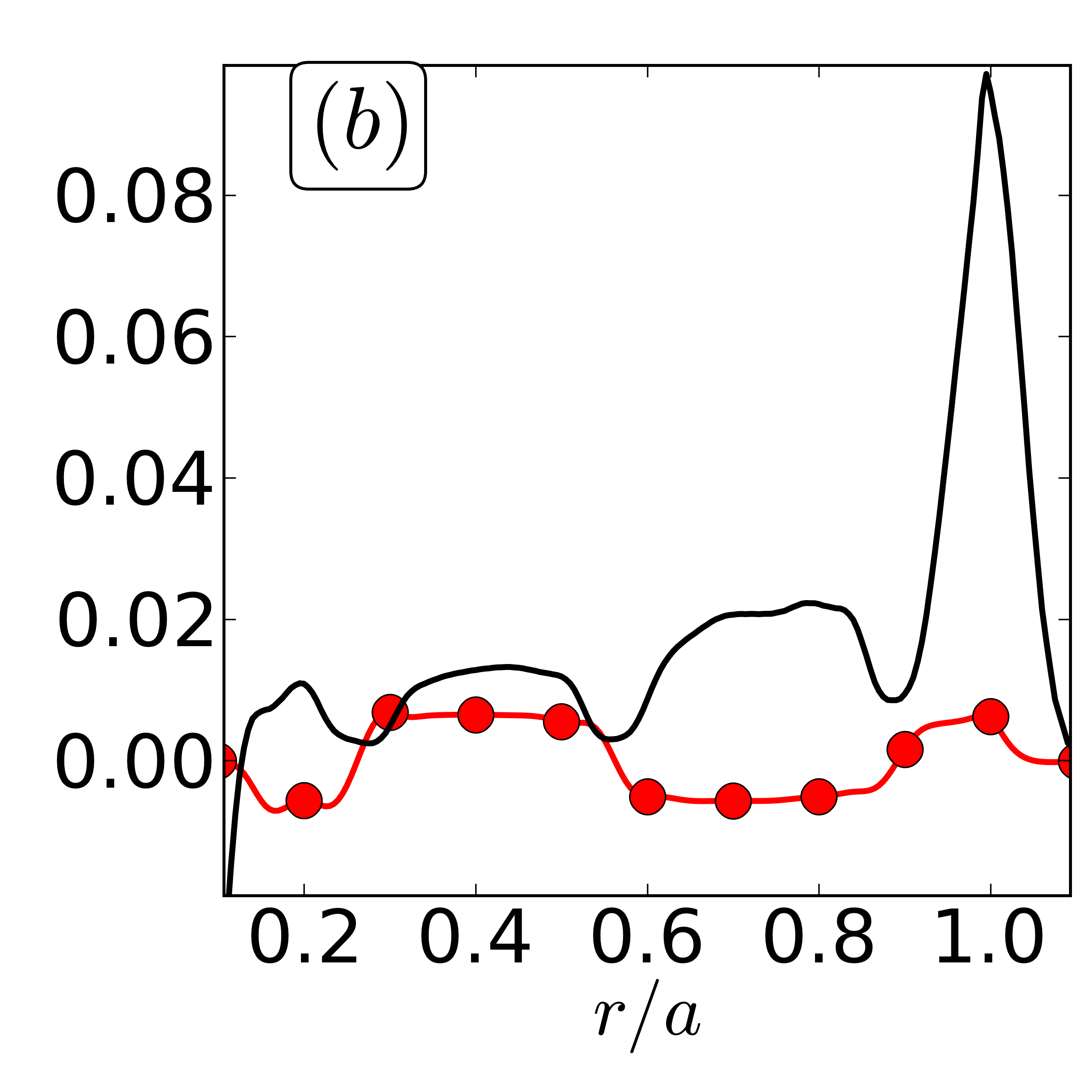}
	\label{subfig:fig2_b}
\end{subfigure}
\end{minipage}
\begin{minipage}{0.2\textwidth}

\begin{subfigure}{\linewidth}
    \caption{}
 	\includegraphics[width=\linewidth]{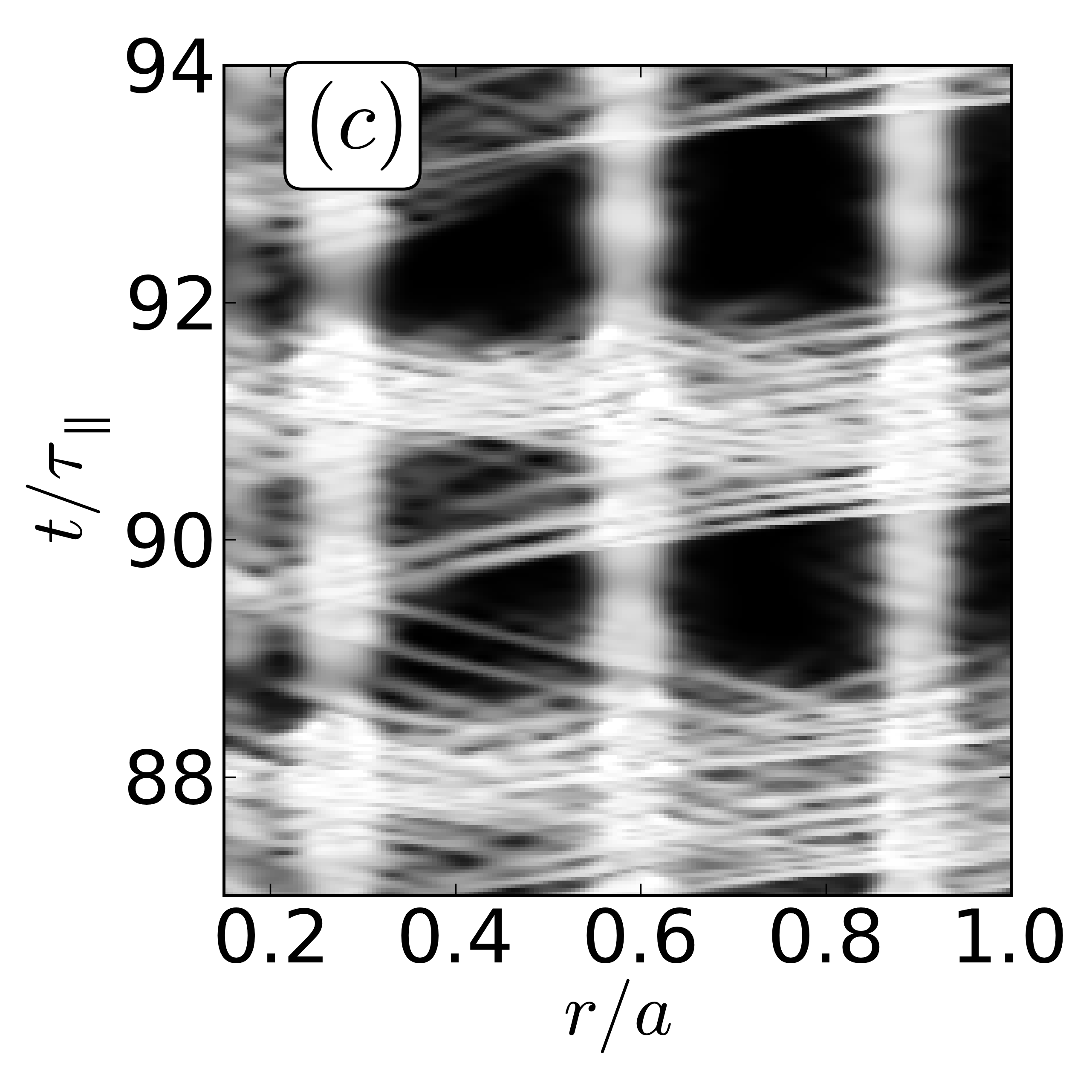}
	\label{subfig:fig2_c}
\end{subfigure}
\vspace*{-1.1cm}

\begin{subfigure}{\linewidth}
    \caption{}
		\includegraphics[width=\linewidth]{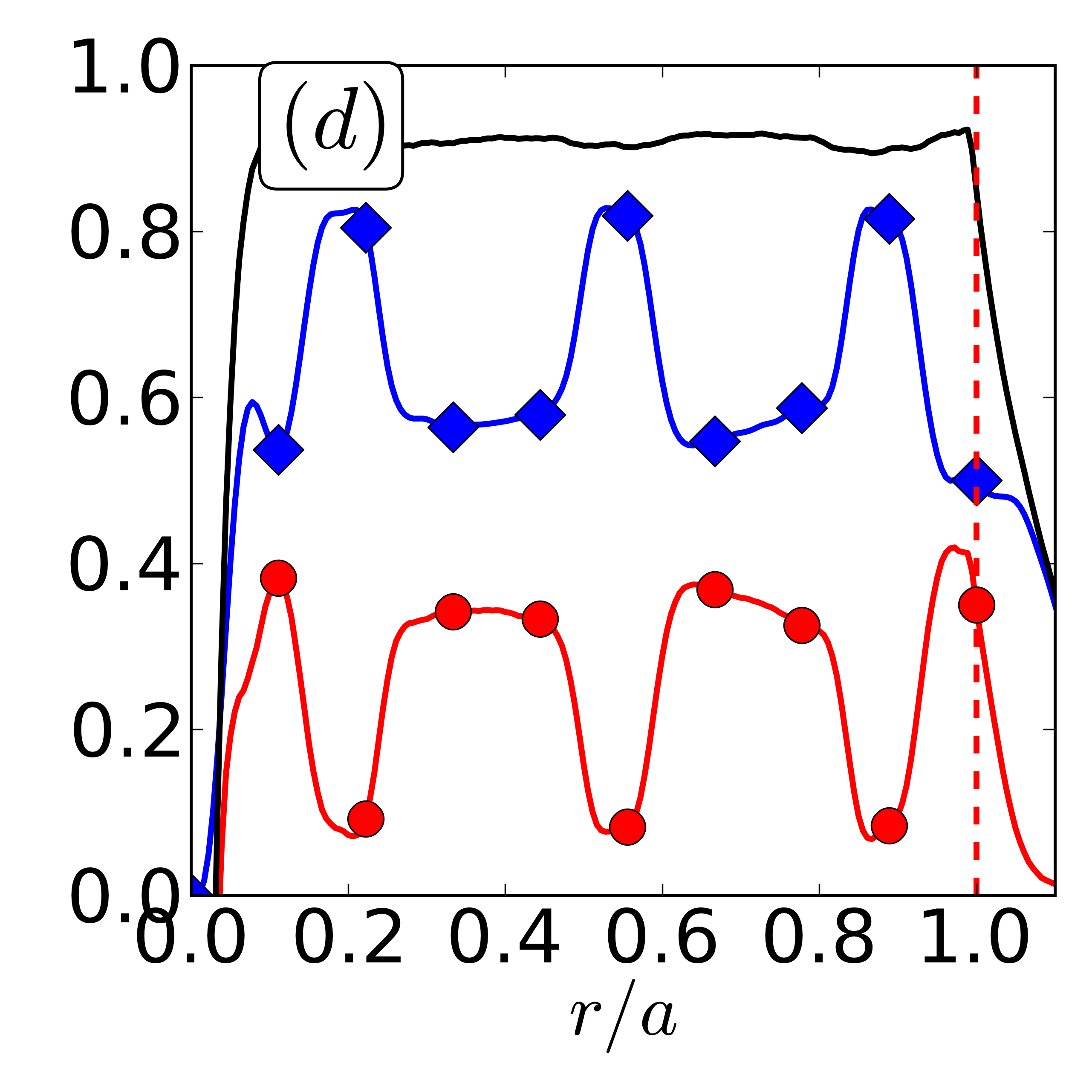}
	\label{subfig:fig2_d}
\end{subfigure}
\end{minipage}
\vspace*{-.6cm}
  \caption{(\protect\subref{subfig:fig2_a}) $E\times B$ shear (red contour) and $1/L_n$ (gray colored) in function of time and radius,(\protect\subref{subfig:fig2_b}) the time and y-averaged shear ( black line)
 and $1/L_n$ (red circles), (\protect\subref{subfig:fig2_c}) parameter $R_b$ used to determine the transport barriers, $R_b$ varies  between 1 (white) and 0 (black), (\protect\subref{subfig:fig2_d}) total flux (black line), turbulent flux (blue diamonds) and diffusive flux (red circles) in the radial direction.}
\label{fig:exb_ln}

\end{figure}

In order to quantify the extent of the barrier in the two regions, we shift the profiles to the same separatrix position and compare the density gradient profiles $1/L_n=-\partial_x n ~/~ n $, fig.\ref{fig:fig1}(\protect\subref{subfig:fig1_b}). The strong density drop observed in the pedestal leads to a marked peak in the $1/L_n$ profile localized at the separatrix. 
The barrier extends into both the edge and SOL region and its width is observed to range between 5 \% and 10 \% of the minor radius  fig.\ref{fig:fig1}(\protect\subref{subfig:fig1_c}).  
Regions with large zonal flows shear are correlated with the corrugation of the profile, see fig.\ref{fig:exb_ln}(\protect\subref{subfig:fig2_a})\&(\protect\subref{subfig:fig2_b}). They are characterized by a stopping capability of most of the avalanches both overdense from uphill and holes from downhill. \\
The simulation is characterized by a slow reorganization of the zonal flow pattern as readily observed on the contour plot of the zonal flow shear superimposed on the 2D plot of $1/L_n$, fig.\ref{fig:exb_ln}(\protect\subref{subfig:fig2_a}). 

\begin{figure}[h]
\vspace*{-.8cm}
  \begin{subfigure}{0.2\textwidth}
    \caption{}

		\includegraphics[width=\linewidth]{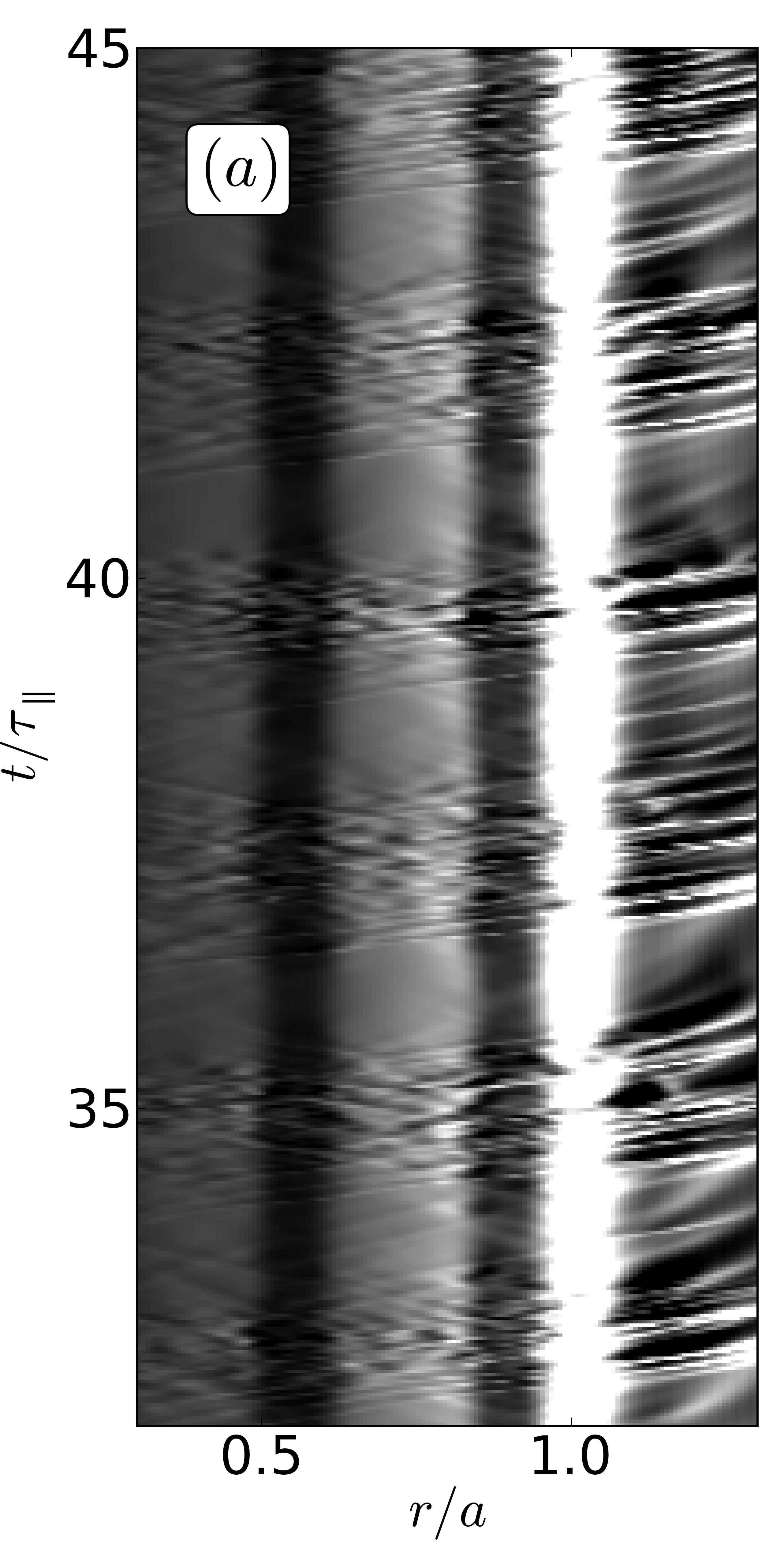}
	\label{subfig:fig3_a}

\end{subfigure}
\hspace*{-.0cm}
\begin{minipage}{0.2\textwidth}
\begin{subfigure}{\linewidth}
    \caption{}

		\includegraphics[width=\linewidth]{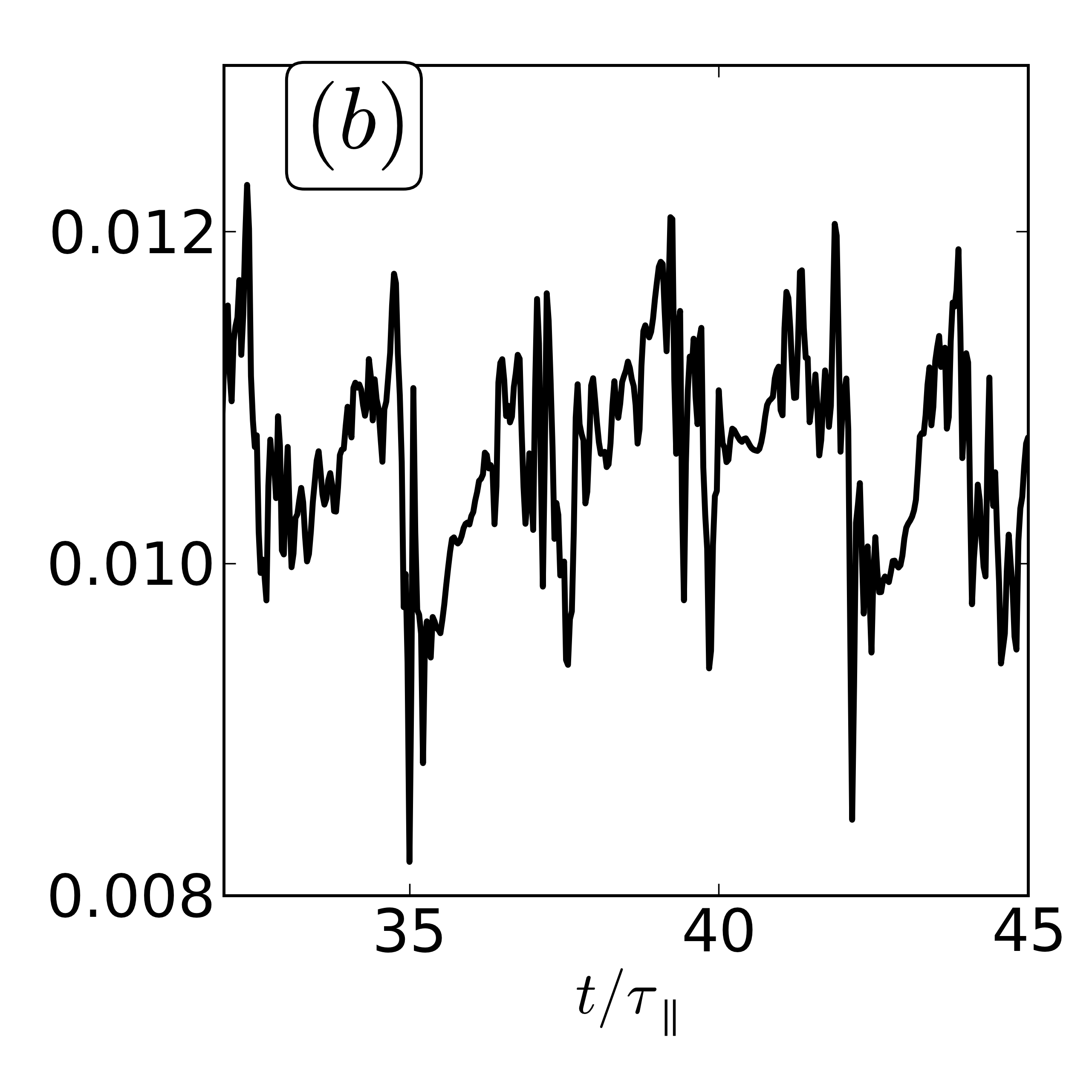}
	\label{subfig:fig3_b}
\vspace*{-1.1cm}

\end{subfigure}
\begin{subfigure}{\linewidth}
    \caption{}

		\includegraphics[width=\linewidth]{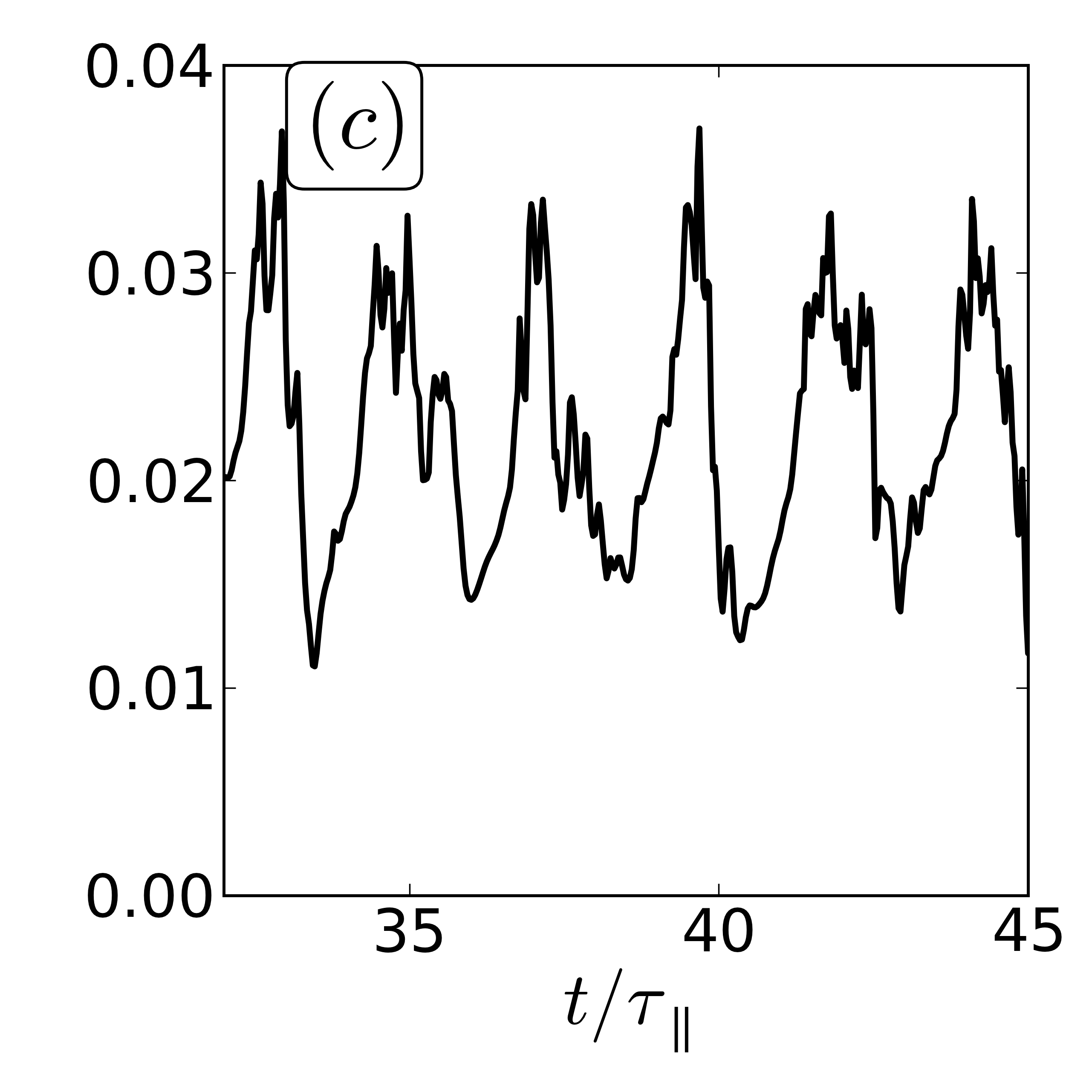}
	\label{subfig:fig3_c}
\end{subfigure}
\end{minipage}

\vspace*{-.5cm}
\caption{(\protect\subref{subfig:fig3_a}) $1/L_n$ evolution in time and radial direction:  the different transport reorganization in SOL and edge region, evolution of  $1/L_n$ in time at given radial positions in the edge (\protect\subref{subfig:fig3_b}), and SOL (\protect\subref{subfig:fig3_c}) }
  \label{fig:sol_edge_tran}
\vspace*{-.3cm}
\end{figure}

Two features are outstanding, the evolution towards a dipolar structure of $<\Phi>_y$ in the edge region while the structure of the maximum value at the separatrix weakly evolves.  When a statistical steady state is reached, one can average the profiles over time, fig.\ref{fig:exb_ln}(\protect\subref{subfig:fig2_b}) and fig.\ref{fig:exb_ln}(\protect\subref{subfig:fig2_d}).
 One can see that the total flux $\Gamma_{tot}$ is radially constant in the edge and decays in the SOL. The turbulent contribution $\Gamma_{turb}$ exhibits well defined minima, and conversely large values of the diffusive flux $\Gamma_{dif}$, of $1/L_n$ and of the zonal flow shear. Narrow regions with strong turbulent transport are localized in the vicinity of the zero zonal flow shear layers. 
To identify the transport barriers, one defines the ratio between the y-averaged particle fluxes $R_b=\Gamma_{turb}/\Gamma_{tot}$ \cite{Floriani2013}.
$R_b$ varies between 0 and 1 in steady state and is a measure of the effectiveness of the barrier in reducing turbulent transport. One readily observes on fig.\ref{fig:exb_ln}(\protect\subref{subfig:fig2_c}) that $R_b$ changes in time and space (x-direction).
  In space, one recovers the dipolar structure with four transport barrier regions, a pedestal at the separatrix that is relatively narrow, two broad barriers in the edge and finally a small transport barrier towards the source region that is strongly linked to the boundary 
conditions of the model. In time, one can observe quasi-periodic relaxation events characterized by strong turbulent transport across all the barriers. While these events are globally quasi-periodic, the detailed time evolution is specific of each event made of consecutive avalanches that do not extend throughout the edge region\cite{Khapko2014,Rolland2011}.
 These events are also correlated with the large transport bursts in the SOL region, see fig.\ref{fig:sol_edge_tran}(\protect\subref{subfig:fig3_a}). As can be observed on the time traces, the edge region exhibits a sawtooth structure corresponding 
to storage by the TB while the SOL region exhibits a pulse-like variation since the SOL acts as the sink for the particles released at each relaxation, fig.\ref{fig:sol_edge_tran}(\protect\subref{subfig:fig3_b})\&(\protect\subref{subfig:fig3_c}).

The generation of a barrier, associated in this model to a corrugation structure in the edge region, is observed for values of the control parameters that are close to the interchange marginal values. The interplay zonal-turbulent modes can be addressed  in the framework of nonlinear
three mode coupling, $\phi_{z}(\kappa,0)$, $\phi_{s}(0,ky)$, $\phi_{t}(\kappa,-ky)$, respectively the zonal flow, a streamer and a more homogeneous turbulent mode \cite{Diamond2005,Kartashova2012}. Setting the interchange instability to zero, $g=0$, we address the dispersion relation in two cases (1) Z-flow generation with growth rate $\gamma_1$,
 finite streamer $\phi_{s}$ mode as reference equilibrium and two coupled modes as perturbations $\phi_{z}$, $\phi_{t}$, conversely,
(2) zonal flow saturation with  growth rate $\gamma_2$, finite zonal equilibrium mode $\phi_{z}$, and perturbations $\phi_{s}$, $\phi_{t}$.
The dispersion relations for case (1) and (2) are:
\begin{subequations}
  \label{eq:3mode coupling}
\begin{align}
  \label{eq:turbulence}
 (\gamma_1+\gamma_{z})(\gamma_1+\gamma_{t})=~~V_t k_y^2\kappa^2|\phi_{s}|^2 \\
\label{eq:streamer} 
(\gamma_2+\gamma_{t})(\gamma_2+\gamma_{s})=- V_{t} k_y^2 \kappa^2 |\phi_{z}|^2
\end{align}
\vspace*{-.1cm}
\end{subequations}
where $\gamma_{z}=\nu \kappa^2$, $\gamma_{s}=\nu k_y^2+\sigma/k_y^2$  and $\gamma_{t}=\nu k^2+\sigma/k^2$, $k^2=\kappa^2 + k_y^2$. The coupling term is $V_t=\left( {k_{y}^2-\kappa^2}\right) /{k^2}$. It governs a symmetric necessary condition for positive growth rates is obtained: in case (a) $k_y^2>\kappa^2$; in case (b)   $k_y^2<\kappa^2$.
\\
Considering the turbulence spectrum in $k_y$, we define three regions: the zonal flow (Z) $k_y=0$, the Big (B) $|k_y|<|\kappa|$ and Small (S) turbulent structures  $|k_y|>|\kappa|$. Region S is the source of zonal flows while B acts as a sink and energy loss via the large scale damping term $\sigma_\phi$. 
The viscosity term  is a generic version of large scale damping, including collisional damping, provided $\kappa\ne0$, that actually controls the linear damping of the zonal flows \cite{Lin1999}. 
The existence of the B-mode region is essential for the transition from a turbulent regime to the zonation regime  characterized by the development of a non homogeneous spectrum, $|k_y| < |k_x| \approx \kappa$.
 Furthermore, the shearing capability of the $k_x=\kappa$ zonal flow is small for the B-modes since $|k_y|<|\kappa|$. For $g\ne0$, the energy injection in the spectrum is governed by the interchange at $k_y\approx \bar{k}$ and the inverse cascade is controlled by the Rhine scale $L_R$, eq.(\ref{eq: growth rate and Rhine scale}). 
The development of the B-mode gap between turbulence and zonal flows is thus constrained by $\kappa_R > \kappa$, where $\kappa_R=\min(\bar{k}, 1/L_R)$.\\
\begin{figure}[h]
\vspace*{-1.cm}
  \begin{subfigure}{0.22\textwidth}
    \caption{}

		\includegraphics[width=\linewidth]{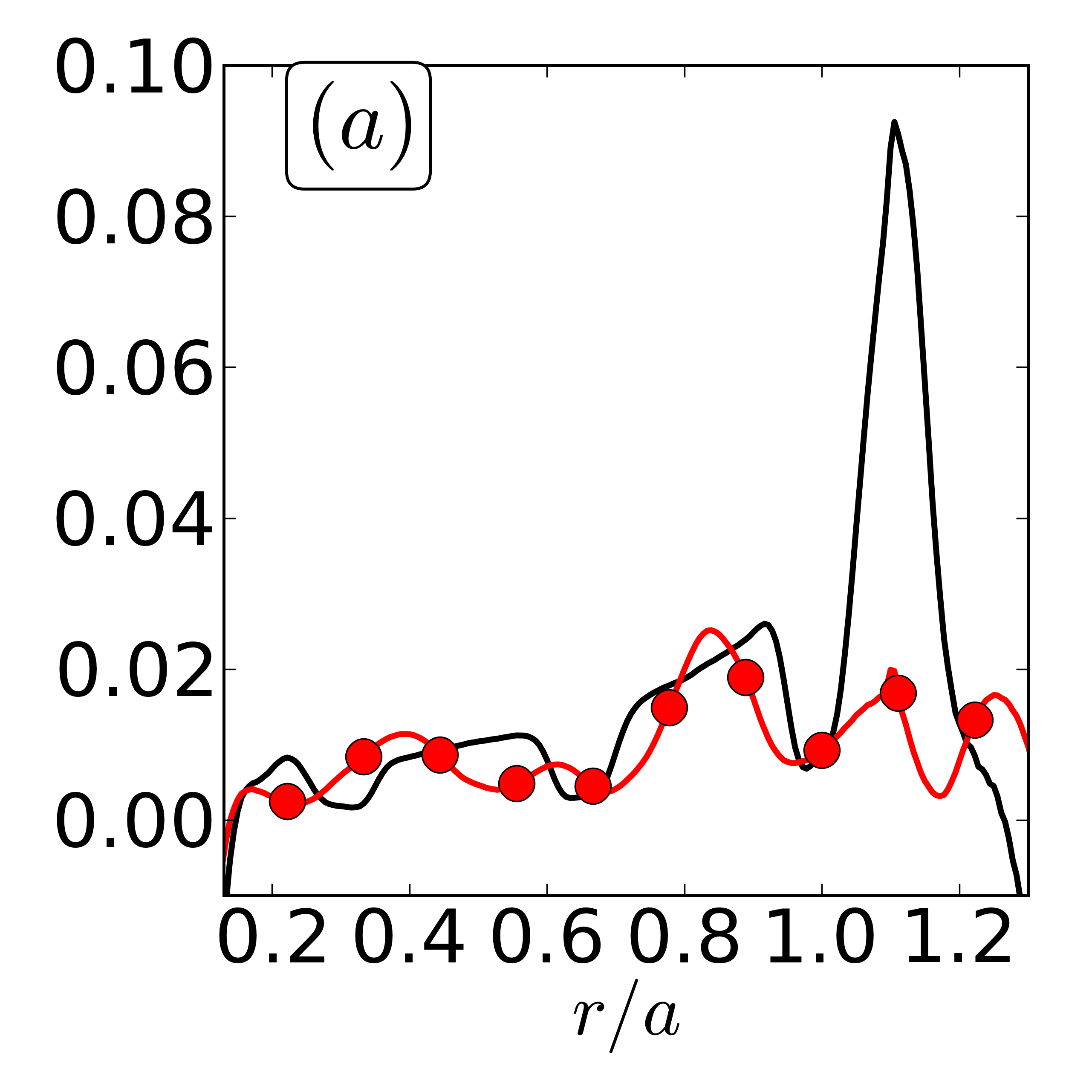}
	\label{subfig:fig4_a}
\end{subfigure}
\hspace*{-.0cm}
\begin{subfigure}{0.22\textwidth}
    \caption{}

		\includegraphics[width=\linewidth]{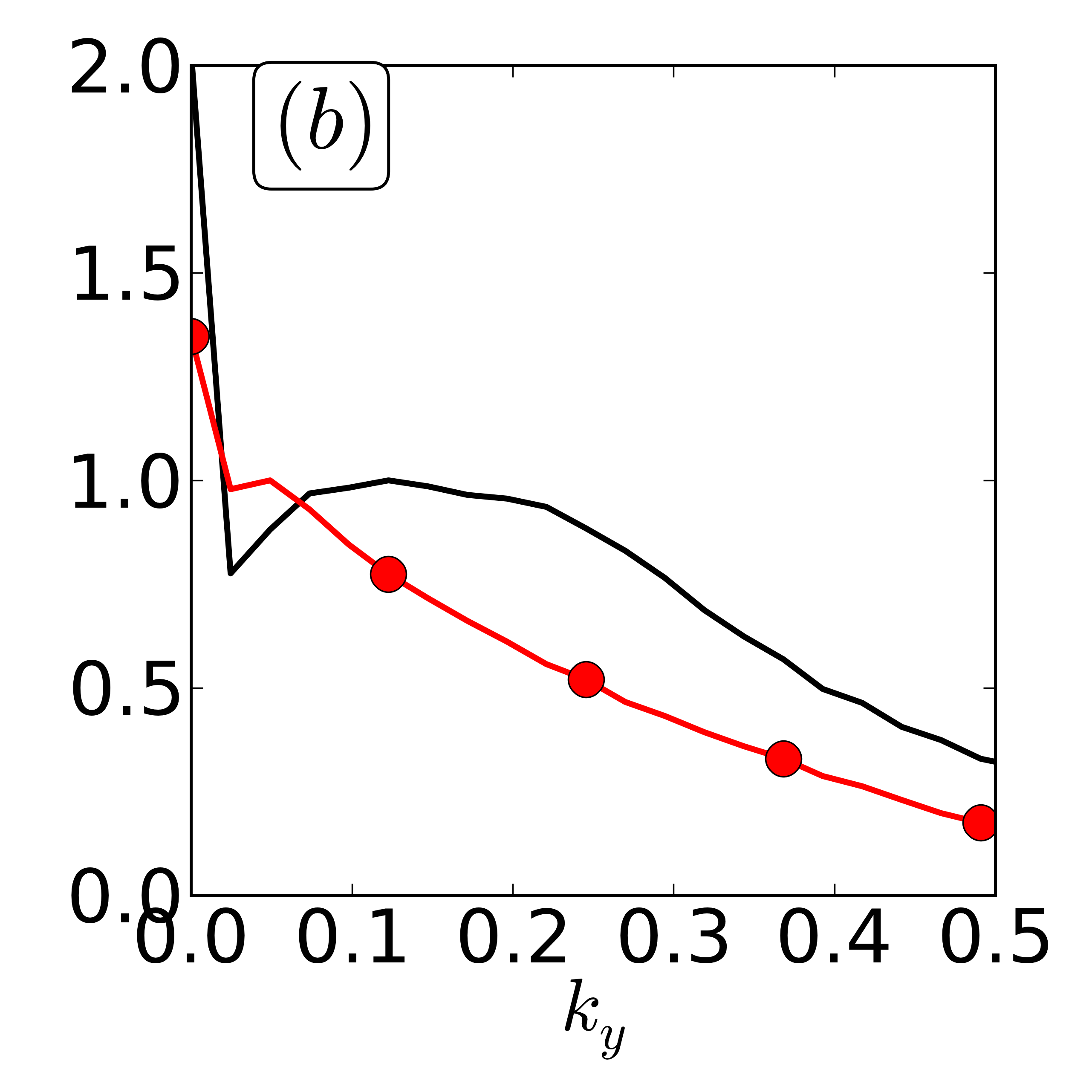}
	\label{subfig:fig4_b}
\end{subfigure}
\vspace*{-.5cm}
\caption{(\protect\subref{subfig:fig4_a}) Density gradient profile and (\protect\subref{subfig:fig4_b}) electric potential spectrum : zonation regime (black line) and turbulent regime (red circles) }
 \label{fig:spectre_r_ky}
\vspace*{-.3cm}
\end{figure} 

\begin{figure}[h]
\vspace*{.2cm}

  \begin{subfigure}{0.22\textwidth}
  		\includegraphics[width=\linewidth]{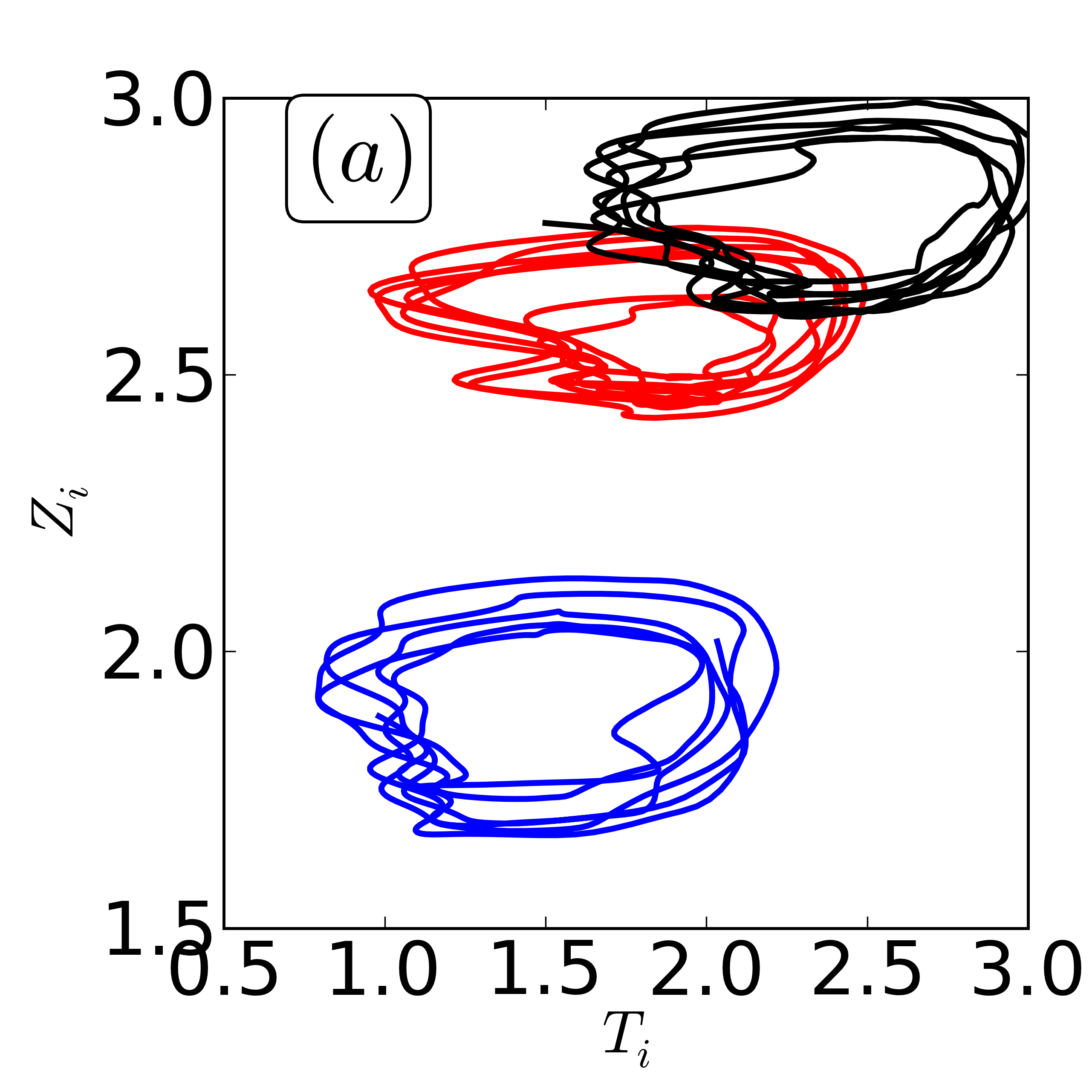}
  \caption{}
	\label{subfig:fig5_a}
\end{subfigure}
  \begin{subfigure}{0.22\textwidth}
		\includegraphics[width=\linewidth]{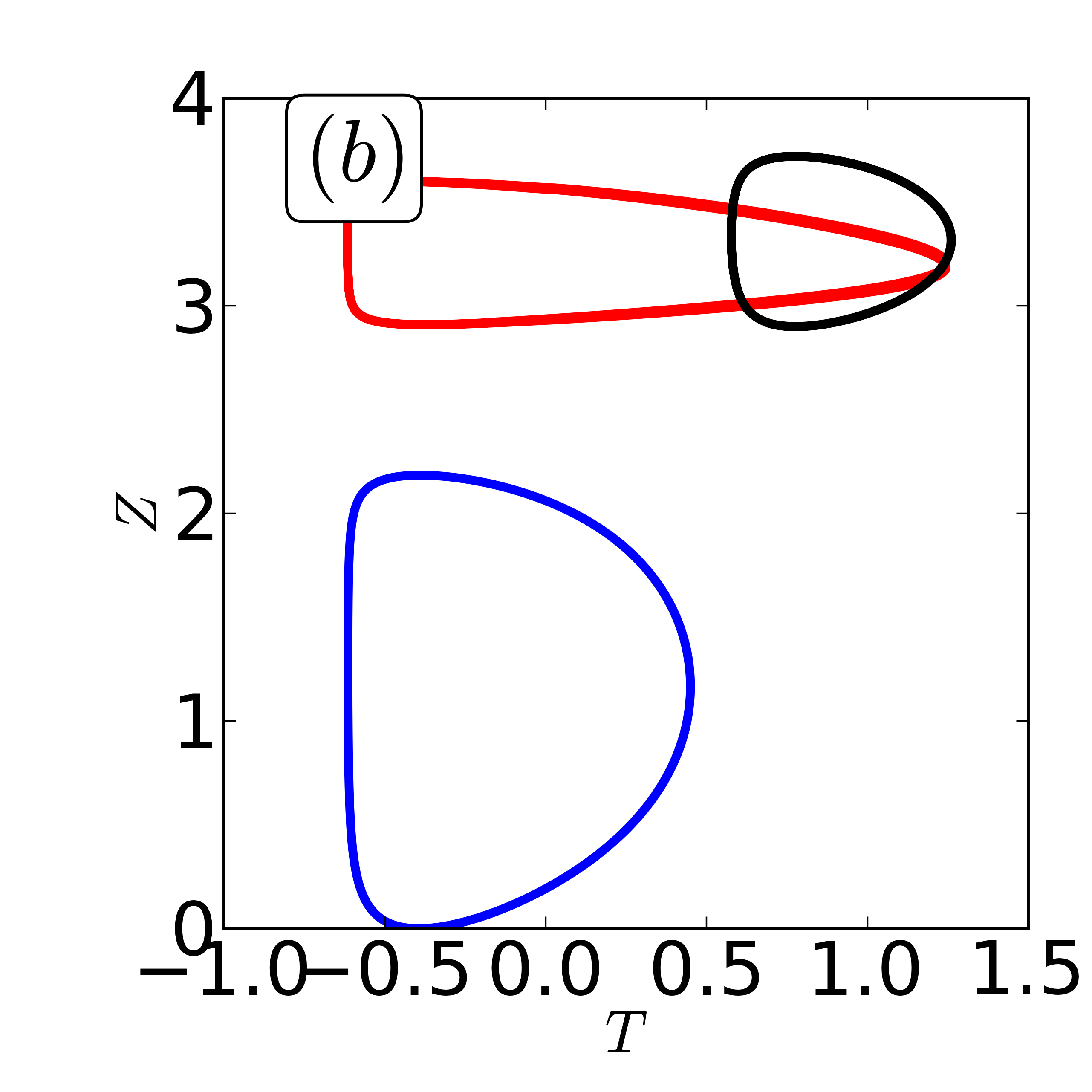}
    \caption{}
	\label{subfig:fig5_b}
\end{subfigure}
\vspace*{-.4cm}
  \caption{Limit cycles: (\protect\subref{subfig:fig5_a}) 2D simulation,(\protect\subref{subfig:fig5_b}) 0D model}
  \label{fig:lr_15}
\vspace*{-.4cm}
\end{figure}

The spectrum of the electric potential in simulations in the zonation regime,  fig.\ref{fig:spectre_r_ky}(\protect\subref{subfig:fig4_b}), is characterized by a gap between S and Z modes, hence with weak B-mode turbulence.
 The S-region transfers energy via non linear coupling towards the Z-flows, which tends to store the energy and quench the turbulent transport dominated by the S-modes. 
The Z-mode is gradually damped by the viscosity damping while the TBs gradients build-up until a new relaxation event is triggered during which the gradients relax and Z-flows are regenerated by the S-mode turbulent activity. 
When the gap between S and Z regions is reduced, hence increasing the B-mode amplitude, the B-mode turbulent activity is not affected by Z-flow shearing and the Z-mode energy can be transferred back to the turbulent modes, both the B and S modes. 
There is then an increase of the frequency of the turbulent bursts to the point where they cannot be isolated from the steady state transport activity and the pedestal is smeared-out,  fig.\ref{fig:spectre_r_ky}(\protect\subref{subfig:fig4_a}). The fluctuations spectra of these two regimes present a strong similarity with the experimental observation achieved during this L-H transition on the stellarator H-1 \cite{Shats2005}. \\
The dynamics of this transition between pedestal (High confinement)  and no-pedestal (Low-confinement) behavior, is captured by the following 0-D predator-prey model \cite{Diamond2005}. 
\begin{subequations}
  \label{eq:predator-prey}
\vspace*{-.3cm}
	\begin{align}
  \frac{\partial_t \nabla n}{\nabla n} = \frac{P}{\nabla n} -\left(  T+T^*\right) 
\\
\frac{\partial_t Z}{Z} = \beta\frac{ S}{T}-\nu
\\
  \frac{\partial_t S}{S} = \gamma_s (\nabla n-\nabla n^*) (1-\alpha_{s}S)-\beta\frac{ Z}{T}
\\
    \frac{\partial_t B}{B} = \gamma_b(\nabla n-\nabla n^*)(1-\alpha_{b}B)
  \end{align}
\end{subequations}
The model couples the gradients $\nabla n$ and the Z, S and B modes. The gradient $\nabla n$ is governed by a balance between the source $P$ and transport, both turbulent $T = B+S$ and collisional $T^*$.
The growth rates for S and B modes are $\gamma_s$ and $\gamma_b$ respectively and exhibit a threshold in the gradient, $\nabla n^*$. Non-linear saturation of these modes are used, proportional to $\alpha_s$ and $\alpha_{b}$. 
The control parameter of Z-flow generation by S is $\beta$. The B-modes act as saturation term on this energy exchange via the $1 / T$ dependence. The Z-flow sink is proportional to the viscosity $\nu$. 
The limit cycles of the Z-T interplay in simulations and 0-D model are compared in the $Z-T$ plane, fig.\ref{fig:lr_15}(\protect\subref{subfig:fig5_a})\&(\protect\subref{subfig:fig5_b}). $Z$ and $T$ are readily determined by eq.(\ref{eq:predator-prey}) for the 0-D model.
 For the simulation output of the interchange turbulence, we define $Z_i = FT(V_z)$ and $T_i^2=FT(R)$, where $FT$ is the 2D Fourier mode energy of the zonal flow velocity $V_z$ and Reynolds stress $R$ respectively.

 The different positions of the cycles in the Z-T plane is determined by the control parameters: from a reference case, blue trace, increasing the curvature term $g$, red trace or decreasing $\sigma$ black trace, fig.\ref{fig:lr_15}(\protect\subref{subfig:fig5_a}).
Increasing $g$ in the simulation leads to an increase of both turbulence and zonal flows, the latter being more important, increase of $Z_i / T_i$.
A comparable behavior is obtained by increasing $\gamma_s$ and $\gamma_b$, namely the growth rate of the interchange instability -governed by $g$- fig.\ref{fig:lr_15}(\protect\subref{subfig:fig5_b}).
Decreasing $\sigma$ governs a decrease of both $\bar{k}$ and $1/L_R$ so that the spectrum maximum shifts towards the low $k_y$ values, reducing the Z-S gap.
The turbulence amplitude is increased as well as the ratio between B and S modes. Consistently, this behavior is recovered in the 0-D model by reducing $\alpha_s$ and $\alpha_b$, the non-linear turbulence saturation, as well as the critical gradient $\nabla n^*$.\\
Barrier generation at the interface between regions with different zonal flow damping is governed by a zonation regime in the plasma edge region. It is triggered by a gap in the turbulent spectrum between the wave vector of the maximum and zero, the zonal flow wave vector. This gap is controlled by the energy injection wave vector of the interchange instability and the Rhine scale that bounds the inverse cascade. Increasing the magnitude of the turbulence drive at given gap reinforces the barrier. In the pedestal High confinement regime, relaxation bursts of turbulence regenerate the zonal flows that are eroded by collisional damping. For fusion plasmas the duration of the quiescent phase between the quasi-periodic relaxation events is thus governed by the ion pedestal collision frequency.\\

The authors thank \mbox{P. Diamond}, \mbox{C. Baudoin}, \mbox{Ch. Passeron} and acknowledge fruitful interactions at the Festival de Th\'eorie, Aix-en-Provence 2013. 
\bibliography{ref_ped2}

\end{document}